\documentclass[18pt a4paper]{article}
\usepackage[T1]{fontenc}
\usepackage[utf8]{inputenc}
\usepackage{authblk}
\usepackage{amsmath,amssymb}
\usepackage{mathtools}
\usepackage{gensymb}
\usepackage{cancel}
\usepackage{booktabs}
\usepackage{graphicx}

\begin{document}

\title{\textbf{Detection of Earth-skimming UHE tau neutrino with the JEM-EUSO detector  }}
\author{Galina Vankova$^1$\thanks{Corresponding author: \texttt{galia@phys.uni-sofia.bg}}, Stefan Mladenov$^1$, Marian Bogomilov$^1$, Roumen Tsenov$^1$, Mario Bertaina$^2$, Andrea Santangelo$^3$}

\date{}
\maketitle
$^1${Department of Atomic Physics, St. Kliment Ohridski University of Sofia}\\
$^2${University of Turin, Italy}\\
$^3${Institut f\"{u}r Astronomie und Astrophysik, Eberhard Karls Universit\"{a}t T\"{u}bingen, Germany} 
\centering
\author{for the JEM-EUSO Collaboration}

\abstract{The ultra high energy cosmic neutrinos are powerful astrophysical probes for both astrophysical
	mechanisms of particle acceleration and fundamental interactions. They open a window into the very distant and high-energy Universe that is difficult to access by any human means and devices. The possibility of detecting them in large exposure space-based apparatus, like JEM-EUSO, is an experimental challenge.
 In this paper we present an estimation of the feasibility of detection of UHE tau neutrino by the JEM-EUSO telescope. The interactions of tau-neutrino in sea water and Earth's crust have been investigated. The estimation of the propagation length and energy of the outgoing tau-lepton shows that if its decay occurs in the atmosphere close enough to the Earth's surface, e.g. below $\sim$ $5 km$ altitude, the cascade is intensive enough and the generated light can be detected from space. We have evaluated the geometrical aperture of the JEM-EUSO detector for the Earth-skimming (horizontal and upward-going) tau-neutrinos by making specific modifications to the standard CORSIKA  code and developing an interface to the existing ESAF (EUSO Simulation and Analysis Framework) software.}

\newpage
\section{Introduction}

The investigation of the ultra high energy cosmic rays (UHECRs) and neutrinos origin is one of the main scientific objectives of the experiments recently searching for high energy cosmic neutrinos such as ANITA \cite{Gorham}, IceCube \cite{Abbasi}, the Pierre Auger Observatory \cite{Abraham}, RICE \cite{Kravchenko} and the forthcoming ARA \cite{Allison}, ARIANNA \cite{Gerhardt}, JEM-EUSO \cite{Takahashi} and LUNASKA \cite{James}.
 Since the attempts of correlating the arrival directions of UHECRs with known active galactic nuclei (AGNs) yield no convincing results \cite{PeirAuger2010,PeirAuger2013} other high energy sources to which the existence of the UHECRs could be connected have been suggested. These include hypernovae (HNe), galactic shocks, gamma-ray bursts, etc.
A possible clue for the sources of interest would be the detection of UHE neutrinos.\\ 
The UHE neutrinos are expected to be born as muon or electron neutrinos. Due to vacuum oscillations, however, the neutrino flux at the Earth is expected to be almost equally distributed among the three neutrino flavours.
 Neutrinos can interact in a given medium either through charged current (CC) or neutral current (NC) channels: $\nu_{l}N \rightarrow l^{\mp}X$ and $\nu_{l}N \rightarrow \nu_{l}X$, where $l = e, \mu, \tau$ is the outgoing lepton and X is the hadron final state. The CC interactions are more important for the space observations because in the NC interactions the secondary neutrino takes away most of the incident energy. Thus, the shower produced by the low energy hadronic component resultant from the NC interaction can not be observed by a detector configuration such as JEM-EUSO, because the shower energy is typically lower than the energy threshold of the detector except for the very high-energy upcoming neutrinos (of the order of $10^{21} eV$).\\ 
Approximately two out of three tau-leptons decay hadronically \cite{Beringer}, Table \ref{tab:taudecay}. From the leptonic channels only $\tau^{-}\rightarrow e^{-}\nu_{e}\nu_{\tau}$ is relevant for tau-lepton detection through initiation of extensive air shower. Having a muon instead of electron in the decay would make the tau-lepton escaping detection due to the large decay length of the muon at the energies considered thus the shower produced cannot be detected from space.
\begin{table}[ht]
\small
\centering
\begin{tabular}{l l c } 
\hline\hline 
$\bf Decay$ $\bf channel$ & $\bf Secondaries$ & $\bf Branching$ $\bf ratio$ $\bf [\%]  $ \\
\hline\hline
$\tau^{-} \rightarrow \pi^{-}$ $\nu_{\tau}$ &  $\pi^{-}$ & 11.8 \\
$\tau^{-} \rightarrow \pi^{-}$ $\pi^{0} \nu_{\tau}$ & $\pi^{-}$, $\pi^{0} \rightarrow 2\gamma$ & 25.8 \\
$\tau^{-} \rightarrow \pi^{-}$ $2\pi^{0} \nu_{\tau}$ & $\pi^{-}$, $2\pi^{0} \rightarrow 4\gamma$ & 10.79 \\ 
$\tau^{-} \rightarrow \pi^{-}$ $\pi^{-} \pi^{+} \nu_{\tau}$ & $2\pi^{-}$, $\pi^{+}$ & 10.0 \\
$\tau^{-} \rightarrow \pi^{-}$ $\pi^{-} \pi^{+}\pi^{0}\nu_{\tau}$ & $2\pi^{-}$, $\pi^{+}$, $\pi^{0} \rightarrow 2\gamma$ &5.18 \\
$\tau^{-} \rightarrow \pi^{-}$ $\pi^{0} \pi^{0}\pi^{0}\nu_{\tau}$ & $2\pi^{-}$, 3$\pi^{0} \rightarrow 6\gamma$  &1.23\\
\hline 
$\bf Total$ $\bf hadronic$&  & 64.8 \\
\hline\hline 
$\tau^{-}\rightarrow e^{-}\overset{-}{\nu_{e}}\nu_{\tau}$ & $e^{-}$ & 17.8 \\
$\tau^{-}\rightarrow \mu^{-}\overset{-}{\nu_{e}}\nu_{\tau}$ & $\mu^{-}$ & 17.4\\
\hline 
$\bf Total$ $\bf leptonic$&  & 35.2 \\
\hline\hline 
\end{tabular}
\caption{Main tau lepton decay channels.}
\label{tab:taudecay}
\end{table}

A cosmic tau-neutrino could interact in the Earth's atmosphere, sea water or Earth's crust. 
It has been shown that the detection of Earth-skimming tau neutrinos in the $E\geq 10^{18} eV$ range is possible due to the fact that the tau-lepton originating in the neutrino CC interaction can decay close to the Earth's surface and the extensive air shower produced is intensive enough to be observed by both fluorescence and Cherenkov detectors \cite{Domokos,Feng,PalomaresRuiz, Kusenko}. 
 In the atmosphere, the neutrino interaction probability increases with the zenith angle. The Earth's curvature and the density of the different atmospheric layers are the key parameters that influence the neutrino induced extensive air shower development. A very high energy tau-neutrino could interact deeply in the atmosphere and, instead of one-peak signature, to initiate the remarkable double-bang event that could be observed by space based telescopes with the JEM-EUSO design parameters.\\ 
 In this work we study, by simulations, another possibility, first pointed out in \cite{PalomaresRuiz}, namely interactions of quasi-horizontal $85^{0}<\theta<90^{0}$ and upward going $90^{0}<\theta<95^{0}$ tau-neutrinos in higher (than air) den   sity media such as Earth's crust and see water. In the simulations we track out the decays of tau-leptons emerging from ocean or Earth's surface in order to estimate the feasibility of detection of such events by the JEM-EUSO telescope.

\subsection{JEM-EUSO telescope}

JEM-EUSO, Extreme Universe Space Observatory (EUSO) attached to the Japanese Experiment Module (JEM) on board the International Space Station (ISS), is a future mission devoted to investigation of extreme energetic cosmic rays.
The JEM-EUSO apparatus is a remote-sensing space instrument \cite{Kajino}, Fig \ref{fig:detector} that will orbit the Earth every $\approx$ 90 minutes on the ISS at an altitude  $H_{ISS}$ $\approx$  400 km. It consists of photon collecting optics \cite{Zuccaro} and focal surface (FS) detector \cite{Kawasaki} together with its electronics \cite{Casolino}. The telescope optics is composed of three double-sided curved circular Fresnel lenses with 2.65 m maximal diameter. 
When high energy cosmic rays interact with the atomic nuclei of air molecules they initiate extensive air
showers. JEM-EUSO will detect the light from isotropic nitrogen fluorescence excited by the extensive air showers and Cherenkov radiation reflected from the earth surface or dense clouds. When the ultraviolet light hits the optics it is focused onto the focal surface which consists of 137 Photo-Detector Modules (PDMs). Each PDM consists of a set of elementary cells containing arrays of Multi-Anode Photo Multiplier Tubes with 64 pixels with a spatial resolution of $0.074 \degree$ each.  Data ftom each pixel, is acquired within 2.5 $\mu s$ time window when the trigger logic (defined according to the hardware requirements) is activated.
The intensity of the observed light depends on the transmittance of the atmosphere, the cloud coverage and the height of the clouds. For precise measurements JEM-EUSO is equipped with an Atmospheric Monitoring system \cite{Neronov} that will characterize
the atmospheric conditions and thus determine the effective observation aperture with high accuracy. The atmospheric monitoring system consists of an infrared camera and a light detection and ranging system. 
The observation area of the earth surface is determined by the projection of the field of view (FoV) of the optics and the area of the focal surface.
The dimensions of the FoV are $\sim64\degree$ and $\sim45\degree$ on the major and minor
axes, respectively. For these axes, the projected lengths on earth surface are
$\sim500$ km and $\sim330$ km, respectively for altitude of the ISS orbit $H_{ISS}$ = 400 km. The effective solid angle $\Omega_{FoV}$ is  $\sim 0.85$ sterad. \\
For the baseline layout of 137 PDMs, the observation area of the apparatus, $A^{nadir}_{obs}$ for nadir mode, is a function of $H_{ISS}$ expressed by:
\begin{figure}[ht]
	\centering
	\includegraphics[width=0.55\textwidth]{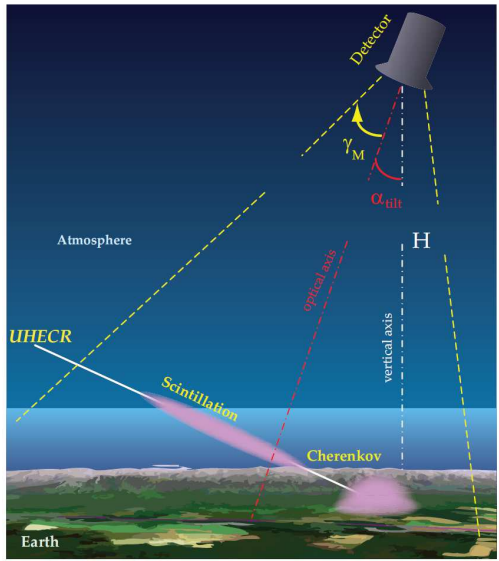}
	\caption{Observational principle of the detector. H is the orbital height, $\gamma_{M}$ the FoV aperture
		semi-angle and $\alpha_{tilt}$ is the tilt angle between the optical axis and the nadir.}
	\label{fig:detector}
\end{figure} 
\begin{equation}
A^{nadir}_{obs} [km^{2}] = \Omega_{FoV}H_{ISS} = 1.4\times10^{5}\Bigl(\frac{H_{ISS}}{400 [km ]}\Bigr)^{2}
\end{equation}

For tilt angles $\alpha < 40\degree$, $A_{obs}(\alpha) = A^{(nadir)}_{obs}(cos (\alpha))^{-b}$,
where, for the altitude of interest b is in the range $3.2 < b < 3.4$. For $\alpha \sim 40\degree - 50\degree$, the detector is sensitive to the area over the local horizon, and $A_{obs}$ saturates above $\alpha \approx 60\degree$. 

\section{Simulation of neutrino interactions in water and earth crust}

\begin{figure}[ht]
\includegraphics[width=1.02\textwidth]{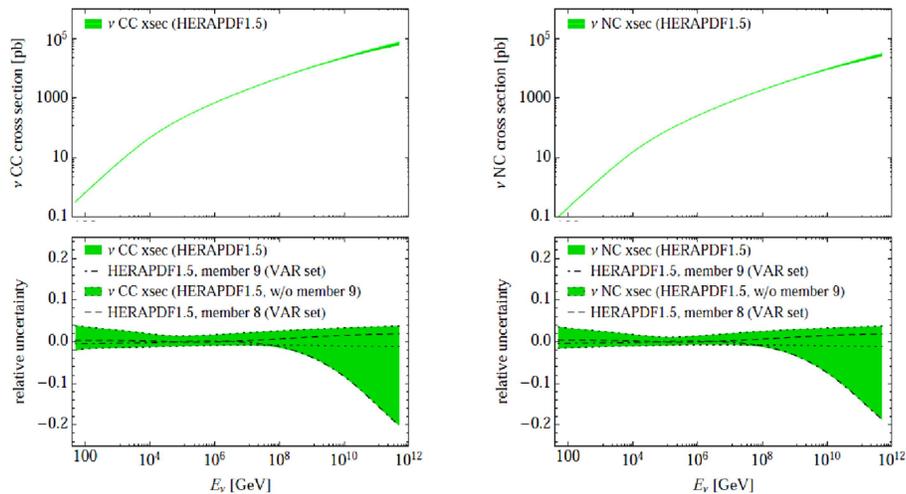}
\caption{Neutrino DIS cross-section for CC and NC scattering as predicted by the HERAPDF1.5. and total relative PDF uncertainties including experimental errors.\protect \cite{Sarkar}}
\label{fig:cross-sect}
\end{figure}

To study the capability of the JEM-EUSO detector to detect UHE neutrino events we have adapted the version 7.401 of CORSIKA code \cite{Heck} with new options for neutrino interactions in water and earth media. For this we have added one layer below the lowest atmospheric one. It corresponds to the last or external layer of the Earth (Earth's crust) with density profile chosen according to the Preliminary Earth Model \cite{Dziewonski} or to a layer with sea water density $\rho_{w} = 1.025 g/cm^{3}$. The user is free to choose the thickness of the layer and therefore to define the interaction volume. 
The observation of tau-neutrinos depends on the incident fluxes (for which a large number of models exists), the cross section and on the detector capability. The major uncertainty on the cross section at the energies we consider comes from the unknown behaviour of the parton distribution functions (PDFs) at very small values of the parton momentum fraction x. For the simulation of the charged and neutral current interactions we have used the deep inelastic neutrino-nucleon cross sections \cite{Sarkar}, see Fig. \ref{fig:cross-sect}, and extrapolated to very small x parton distribution functions CT10 \cite{Sarkar,Thorne1,Thorne2}.
The high energy cross-section is evaluated specifically for experiments devoted to the UHE neutrino measurements. The input PDFs considered treat heavy quarks by using general-mass-variable-flavour number schemes.\\
For primary neutrinos CORSIKA calls the HERWIG Monte Carlo generator \cite{Herwig} to produce the first interaction and then tracks the products of this interaction. The call to HERWIG has been adapted to the cases of water and earth taking into account the dependence of the interactions on the inelasticity parameter $y = 1 - \frac{E_{\tau}}{E_{\nu_{\tau}}}$.
 In absence of reliable predictions for the shape of the UHE neutrino spectrum, we consider
power law fluxes $F_{\nu_{\tau}}$ $\approx$ $E^{-\gamma}$, with $\gamma = 2$ to study how the propagation of tau is influenced by its energy losses.     
\newpage
\subsection{Tau lepton propagation and energy losses}
\begin{figure}[ht]
	\makebox[\textwidth]{
		\includegraphics[width=0.50\textwidth]{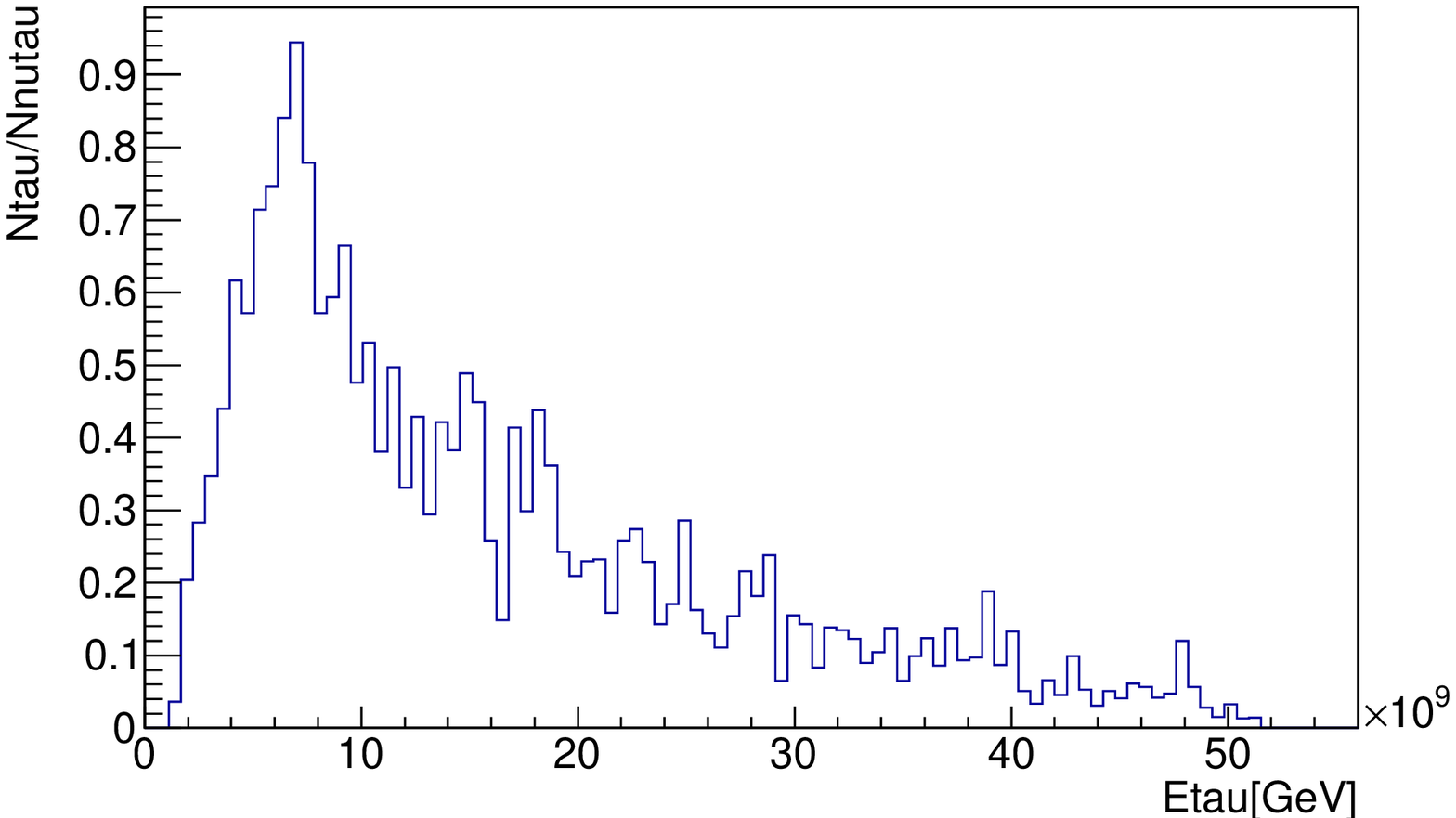}
		\hfill    
		\includegraphics[width=0.50\textwidth]{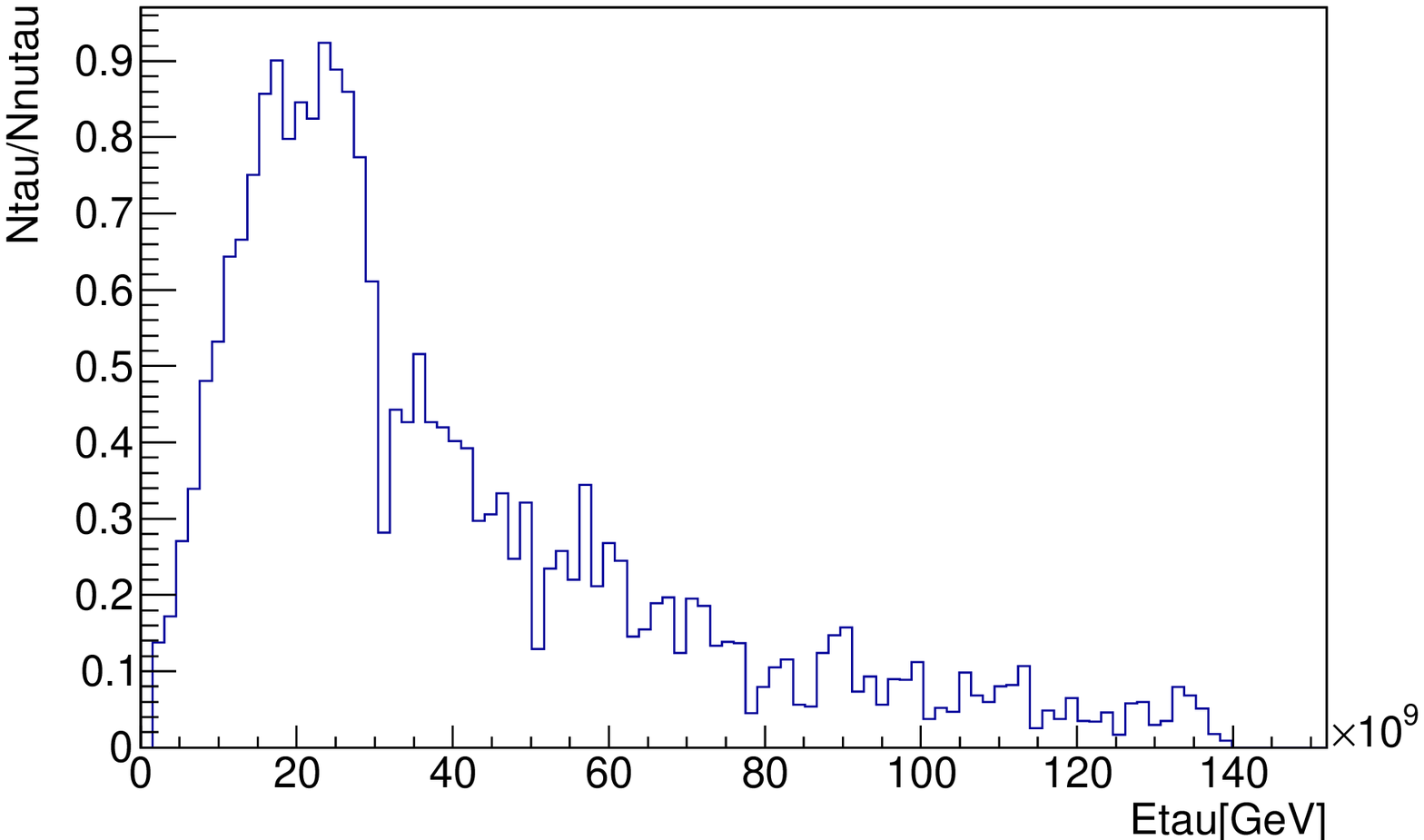}  
	}
	\caption{Energy distribution of tau-leptons emerging from earth (left) and water (right) surfaces normalized to the number of simulated events.}
	\label{fig:Etau}
\end{figure}

The average lepton energy loss per distance  X $[g/cm^{2}]$ traveled is a sum of nearly constant ionization energy loss and weakly energy dependent radiative energy loss through bremsstrahlung, pair production and photo-nuclear scattering \cite{Dutta2001,Dutta2005,Bugaev2004,Bugaev2009,Aramo}. For the case of earth medium and energy region $E > 10^{19} eV$ the dominating photo-nuclear process is responsible for the most uncertain contribution. In our code the energy losses of the charged leptons in water are computed by a modified stopping power formula to account for the density effect.
The cross sections for tau-lepton interactions with mater we use are taken from \cite{Dutta2001,Bugaev2004, Bugaev2009}. In these cross sections the nuclear structure of the target (elastic and inelastic form factors), screening (shadowing) effects are taken into account. Following the respective CORSIKA procedures we treat the tau propagation stochastically, with a series of electromagnetic interactions occurring according to the cross section and decay probability.
We have simulated $10^{5}$ tau-neutrino events in both media with energies $E_{\nu_{\tau}}$ in the range $10^{19} - 10^{21} eV$.
Our simulations show that the tau-lepton, produced by CC $\nu_{\tau}$ interactions under the surface of the Earth or the ocean emerges in the atmosphere with energies shown in Fig. \ref{fig:Etau}. 

\begin{figure}[ht]
	\centering
	\includegraphics[width=0.50\textwidth]{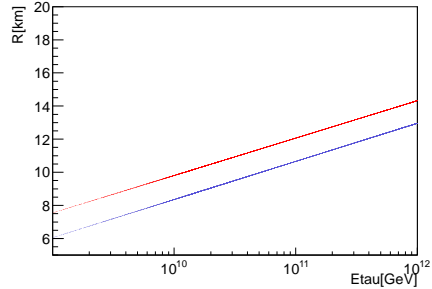}
	\caption{Distance R (propagation length) that the tau-lepton travels in earth (blue line) and water (red line) medium after its production as function of incident neutrino energy.}
	\label{fig:Range}
\end{figure}

The distance R (propagation length) that tau travels after its production in earth or water medium as a function of the incident neutrino energy is shown in Fig. \ref{fig:Range}. For energies above the JEM-EUSO threshold of $3\times 10^{19} eV$ its propagation length ranges from  ${\sim}10$ to ${\sim}15$ km. 
Our results are in a good agreement with the results of \cite{Dutta2005, Fargion2006, Gora2007} despite the difference in methods for calculation of R that have been used in those works.

 The number of showers produced, for tau-neutrino energies above $E = 3\times10^{19} eV$, is shown in Table \ref{tab:numshowers}. As one can see the number of showers decreases with increasing of the primary neutrino energy. For a shower that provides a detectable signal on the focal surface of the detector the energy of its charged component has to be larger than $E_{charged} \sim 2.5 \times 10^{19} eV$. To test the JEM-EUSO sensitivity to the skimming events we have analyzed all the showers with energy of the charged component larger than $2 \times 10^{19} eV$. 
The results indicate that if the decay of the emerging tau-lepton occurs in the atmosphere close enough to the Earth's surface, e.g. below $\sim 5 km$ altitude, the cascade is intensive enough and the generated light can be detected from space. 
The number of showers passing the selection criterion is presented in square brackets, Table \ref{tab:numshowers}.

\begin{table}
	\small
        \centering
	\begin{tabular}{ c c c c c c c}
		\hline\hline
		$ \bf E_{\nu_{\tau}}$ $\times10^{19}$ $[eV]$ & $3 - 5 $ & $5 - 7$  & $7 - 9 $ & $ 9 - 11 $ & $ 11 - 13 $ \\
		\hline\hline
		$\bf interaction$ $ \bf in$ $ \bf earth$  \\
		\hline
		$\theta_{z}=[85^{0}-90^{0}]$ & 2230 $[895]$ & 1012 $[689]$ & 629 $[99]$ & 493 $[49]$ & 415 $[0]$ \\
		$\theta_{z}=[90^{0}-95^{0}]$ & 3864 $[741]$ & 1484 $[663]$ & 856 $[102]$& 615 $[52]$ & 507 $[1]$ \\
		\hline\hline
		$\bf interaction$ $\bf in$ $\bf water$  \\
		\hline
		$\theta_{z}=[85^{0}-90^{0}]$ & 2291 $[1017]$ & 1315 $[689]$ & 714 $[102]$ & 513 $[54]$ & 442 $[1]$ \\
		$\theta_{z}=[90^{0}-95^{0}]$ & 3786 $[984]$ & 1341 $[663]$ & 805 $[106]$ & 621 $[58]$ & 512 $[2]$ \\
		\hline\hline
	\end{tabular}
	\caption{Number of showers produced in the atmosphere for tau-neutrino energies above $E = 3\times10^{19} eV$. The number of the initial interactions is $10^5$ and of the emerged tau leptons is $\sim$ $25\times10^3$. In brackets:  A selection of the showers generated from tau decays occurred below H = 6 km and 
$E_{charged} > 2 \times 10^{19}$ has been applied.}
	\label{tab:numshowers}
\end{table}
\section{Estimation of the geometrical aperture}

\begin{figure}[ht]
	\makebox[\textwidth]{
		\includegraphics[width=0.50\textwidth]{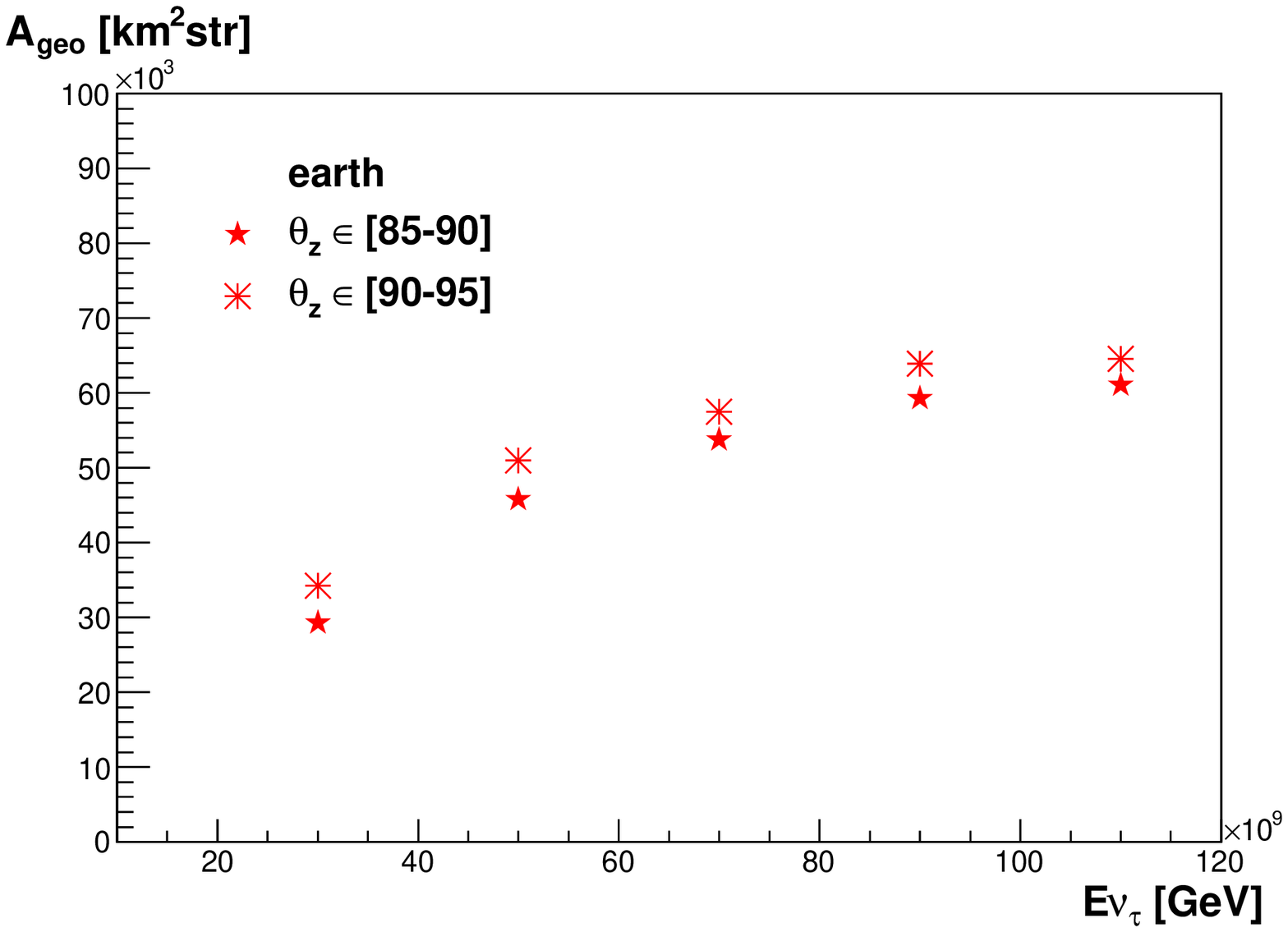}
		\hfill    
		\includegraphics[width=0.50\textwidth]{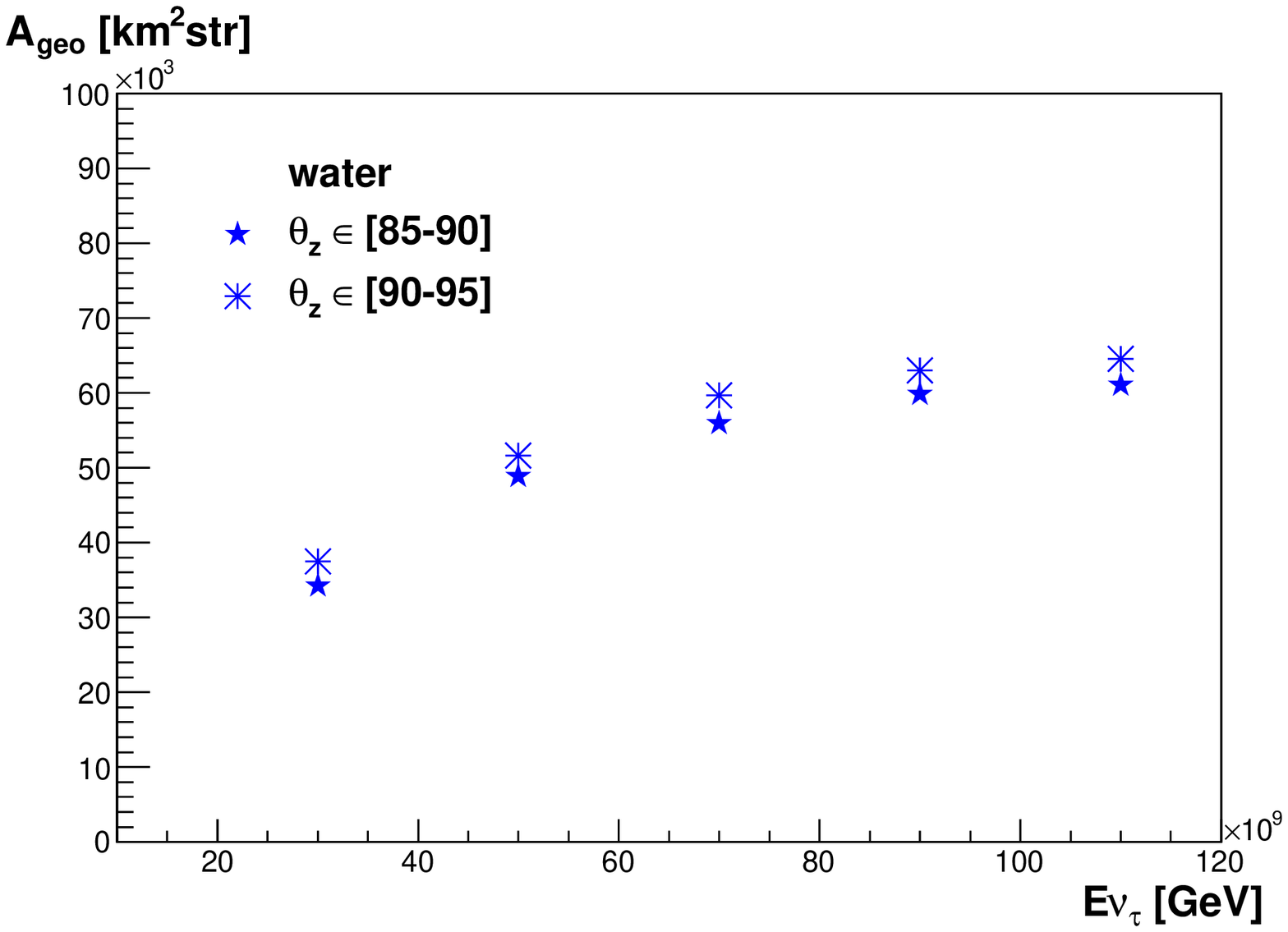}
	}
	\caption{Geometrical aperture of the JEM-EUSO for horizontal and upward going neutrinos interacted in earth (left) and water (right) medium.}
	\label{fig:geomaperture}
\end{figure}

The estimation of the aperture has been done by means of ESAF \cite{Berat} software and a special interface of this code to the CORSIKA package. Both software parts are written in C++ and the output is visualized by the ROOT package \cite{Brun}.  
In our analysis we assume  clear sky condition with average background. By rescaling measurements of the Tatiana satellite \cite{Garipov} to the ISS orbit \cite{Adams,Bobik} the average background flux is estimated to be $\sim 500 $ {\it photons.}$m^{-2}$.$sterad^{-1}$.$ns^{-1}$.\\
The acceptance of the detector is: 

\begin{equation}
	Acc = \int_{s}\int_{\Omega}dS d\Omega\epsilon(\overrightarrow{r},\theta,E,\kappa)cos\theta, 
\end{equation}
where $\epsilon(\overrightarrow{r},\theta,E,\kappa)$ is the detection efficiency depending on
the EAS position $\overrightarrow{r}$, zenith angle $\theta$, energy E and all possible causes of a lack of detection that we denote by $\kappa$. 
We determine the geometrical aperture through the probability of satisfying the second level trigger (Linear Track Trigger) \cite{Bayer} as obtained from our simulation. The trigger follows the movement of the shower spot in the focal surface over a specific time window to distinguish the event pattern from the background.
To take into account also the showers crossing the field of view with a core location out of the observation area, we have injected a number of showers, $N_{inj}$, initiated by tau-leptons with energies shown in Fig. \ref{fig:Etau} and emerging angles close to $90^{0}$ in an effective area $S_{inj}$ bigger than the observation area of the telescope.\\
For the present JEM-EUSO baseline configuration the geometrical aperture of JEM-EUSO as a function ot the primary energy is:

\begin{equation}
	A_{g}(E_{\nu_{\tau}})=\frac{N_{trigg}}{N_{inj}} . S_{inj} .\Omega_{0}
\end{equation}

where the ratio $\frac{N_{trigg}}{N_{inj}}$ gives the probability of a trigger signal to be created when a shower location $\overrightarrow{r}$ is inside the detector field-of-view.  $\Omega_{0}$ is the effective solid angle of the detector for the zenith angle range of interest. $N_{inj}$ is the injected number of showers initiated in the atmosphere by the tau-leptons emerging from the water or earth medium and $N_{trigg}$ is the number of the events selected by the trigger logic mentioned above.\\ 
Fig. \ref{fig:geomaperture} shows the geometrical aperture as a function of the primary energy for the case of horizontal and upward going neutrinos.

For both media the geometrical aperture increases with the neutrino energy. In case of interactions in the air the dependence of the geometrical aperture on the neutrino energy is approximately the same but it follows the trend until higher energies. It is due to the smaller energy losses of tau and the Earth's curvature; both influence the propagation length of the lepton and the longitudinal particle and energy distributions of the showers produced. Estimation of the detector acceptance and expected event rate for tau neutrino interactions in the three media is subject of a forthcoming paper.
\section{Conclusion}
\linespread{1.3}
A tool for simulation of UHE cosmic tau-neutrino interactions in water and Earth's crust has been developed. The propagation length of the produced tau-leptons in both media as well as their 
energy distributions are presented. The simulation of the detector response to the fluorescent and Cherenkov components of the extensive air showers has been performed by the ESAF package with the interface to the CORSIKA software that we have developed. The results shown in Fig. 3 indicate that the trigger efficiency of the JEM-EUSO telescope increases up to $\sim 100 \%$ at energies $E_{\nu_{\tau}} \geq 8 \times 10^{19} eV $. The geometrical aperture of JEM-EUSO detector for Earth-skimming tau-neutrino events has been estimated for clear sky condition, nadir mode and average background. 
The detected events at energies close to the detector threshold will make possible a cross-check based on the obtained energy spectrum with the ground-based experiments.   
The detailed simulation analysis of the event rate per year and full operational time for tau-neutrino interactions in atmosphere, sea water and Earth's crust is in progress.\\
\section{Acknowledgments}
We express our gratitude to the Research fund of Sofia University for the support of this work through Grant Agreement FNI-SU 53/3.4.15.
This work was supported under the ESA Topical Team Contract No. 4000103396 and by the Bundesministerium für Wirtschaft und Technologie
through the Deutsches Zentrum für Luft- und Raumfahrt e.V. (DLR) under the grant number FKZ 50 QT 1101.

\end{document}